\documentclass[a4paper,11pt]{article}
\pdfoutput=1
\usepackage{jcappub}
\usepackage[T1]{fontenc}
\usepackage{verbatim}
\usepackage{bbm}

\newcommand{\mathd}{\mathrm{d}}
\usepackage{url}

\title{\boldmath Thermalization time scales for WIMP capture by the Sun in effective theories}

\author[a,b]{A. Widmark}
\affiliation[a]{The Oskar Klein Centre for Cosmoparticle Physics, AlbaNova, \\SE-106 91 Stockholm, Sweden}
\affiliation[b]{Department of Physics, AlbaNova, Stockholm University,\\SE-106 91, Stockholm, Sweden}
\emailAdd{axel.widmark@fysik.su.se}

\abstract{I study the process of dark matter capture by the Sun, under the assumption of a Weakly Interacting Massive Particle (WIMP), in the framework of non-relativistic effective field theory. Hypothetically, WIMPs from the galactic halo can scatter against atomic nuclei in the solar interior, settle to thermal equilibrium with the solar core and annihilate to produce an observable flux of neutrinos. In particular, I examine the thermalization process using Monte-Carlo integration of WIMP trajectories. I consider WIMPs in a mass range of 10--1000 GeV and WIMP-nucleon interaction operators with different dependence on spin and transferred momentum. I find that the density profiles of captured WIMPs are in accordance with a thermal profile described by the Sun's gravitational potential and core temperature. Depending on the operator that governs the interaction, the majority of the thermalization time is spent in either the solar interior or exterior. If normalizing the WIMP-nuclei interaction strength to a specific capture rate, I find that the thermalization time differs at most by 3 orders of magnitude between operators. In most cases of interest, the thermalization time is many orders of magnitude shorter than the age of the solar system.}

\begin{document}
\maketitle
\flushbottom

\section{Introduction}\label{section:introduction}
Research in recent decades indicate that non-baryonic dark matter constitutes a majority of the Universe's matter content. This is supported by observational evidence of a vast physical range, from the sub-galactic to the cosmological scale \cite{JungmanKamionkowskiGriest96,Bergström00,BertoneHooperSilk05}. One of the most prominent and studied dark matter particle candidates is the Weakly Interacting Massive Particle (WIMP), for which many different detection techniques are utilized \cite{KlasenPohlSigl15,UndagoitiaRauch16,Gaskins16}.

Indirect detection aims to infer the existence of dark matter particles by observing their annihilation products. A variety of such experiments are conducted, observing various Standard Model particles from various sources \cite{Sushch16,Abeysekara14,Atwood09,Picozza07,Pohl14}. The focus of this paper is the hypothesized process of dark matter capture by the Sun. WIMPs from the galactic dark matter halo can scatter against atomic nuclei in the Sun's interior and become gravitationally bound. With further collisions they lose energy, thermalize and settle in the Sun's core. Annihilation of captured WIMPs could potentially produce a detectable flux of neutrinos emanating from the Sun, differentiable from neutrinos produced by nuclear fusion due to their higher energy scale \cite{SilkOliveSrednicki85,PressSpergel85}. Such a neutrino signal is currently sought after with neutrino telescopes, such as IceCube, Super-Kamiokande, ANTARES and Baikal \cite{Aartsen16,Choi15,Adrian-Martinez16,Avrorin15}.

In recent years, effort has gone into studying dark matter experiments in the framework of non-relativistic effective field theories \cite{FanReeceWang10,Menendez12,Fitzpatrick12,Fitzpatrick13,Fitzpatrick14,Chang10,Gazda16,Barello14,Hill14,Klos13,Peter14,Schneck15,Vietze15}. This allows for a model independent analysis of observational data. The same approach has also been applied to the process of dark matter capture by the Sun \cite{Liang14,Guo14,Blumenthal15,Catena15,CatenaSchwabe15,CatenaWidmark16}, for which capture rates of different WIMP-nucleon interaction operators and WIMP self-interaction operators have been calculated.

The aim of this article is to investigate the process of thermalization for WIMP capture by the Sun, and to do so in the framework of a non-relativistic effective field theory. The reasons for studying this subject are the following. The time it takes for a WIMP bound in orbit to be down-scattered and thermalized to core temperature is most often neglected and approximated as instantaneous. Furthermore, the resulting WIMP density distribution is assumed to follow a thermal profile, given by the Sun's gravitational potential and core temperature. Any departure from these statements can significantly alter the distribution of WIMPs inside the Sun, which in turn has a direct effect on the rate of annihilation and the resulting neutrino signal. There is also a possibility that the WIMP capture rate varies with time, as an effect of the Sun traveling through substructures in the dark matter halo \cite{Koushiappas09}. In such a scenario, a long thermalization time would serve to smoothen fluctuations in the annihilation rate and neutrino signal.

This paper is structured as follows. In section \ref{section:EFT}, I present a brief theoretical summary of WIMP-nucleus scattering in effective field theory. In section \ref{section:capture}, I review the theory of WIMP capture by the Sun, including the method used for simulating the down-scattering process. In sections \ref{section:results} and \ref{section:conclusions}, I present the results and conclusions.
\section{WIMP scattering in non-relativistic effective theories}\label{section:EFT}

\begin{table}
\begin{center}
\begin{tabular}{l @{\hskip 30pt} l}
    \hline\\[-0.4cm]
    $\hat{\mathcal{O}}_1 = \mathbbm{1}_{\chi N}$
    & $\hat{\mathcal{O}}_9 = i\mathbf{\hat{S}}_\chi\cdot\left( \mathbf{\hat{S}}_N\times\frac{\mathbf{\hat{q}}}{m_N} \right)$  \\
    $\hat{\mathcal{O}}_3 = i\mathbf{\hat{S}}_N\cdot\left( \frac{\mathbf{\hat{q}}}{m_N}\times\mathbf{\hat{v}}^\perp \right)$
    & $\hat{\mathcal{O}}_{10} = i\mathbf{\hat{S}}_N\cdot\frac{\mathbf{\hat{q}}}{m_N}$  \\
    $\hat{\mathcal{O}}_4 = \mathbf{\hat{S}}_\chi\cdot\mathbf{\hat{S}}_N$
    & $\hat{\mathcal{O}}_{11} = i\mathbf{\hat{S}}_\chi\cdot\frac{\mathbf{\hat{q}}}{m_N}$  \\
    $\hat{\mathcal{O}}_5 = i\mathbf{\hat{S}}_\chi\cdot\left( \frac{\mathbf{\hat{q}}}{m_N}\times\mathbf{\hat{v}}^\perp \right)$
    & $\hat{\mathcal{O}}_{12} = \mathbf{\hat{S}}_\chi\cdot\left( \mathbf{\hat{S}}_N\times\mathbf{\hat{v}}^\perp \right)$  \\
    $\hat{\mathcal{O}}_6=\left(\mathbf{\hat{S}}_\chi\cdot\frac{\mathbf{\hat{q}}}{m_N}\right)\left(\mathbf{\hat{S}}_N\cdot\frac{\mathbf{\hat{q}}}{m_N}\right)$
    & $\hat{\mathcal{O}}_{13} = i\left(\mathbf{\hat{S}}_\chi\cdot\mathbf{\hat{v}}^\perp\right)\left( \mathbf{\hat{S}}_N\cdot\frac{\mathbf{\hat{q}}}{m_N} \right)$  \\
    $\hat{\mathcal{O}}_7 = \mathbf{\hat{S}}_N\cdot\mathbf{\hat{v}}^\perp$
    & $\hat{\mathcal{O}}_{14} = i\left(\mathbf{\hat{S}}_\chi\cdot\frac{\mathbf{\hat{q}}}{m_N}\right)\left( \mathbf{\hat{S}}_N\cdot\mathbf{\hat{v}}^\perp \right)$  \\
    $\hat{\mathcal{O}}_8 = \mathbf{\hat{S}}_\chi\cdot\mathbf{\hat{v}}^\perp$
    & $\hat{\mathcal{O}}_{15} = -\left(\mathbf{\hat{S}}_\chi\cdot\frac{\mathbf{\hat{q}}}{m_N}\right)\left[\left(\mathbf{\hat{S}}_N\times\mathbf{\hat{v}}^\perp\right)\cdot\frac{\mathbf{\hat{q}}}{m_N}\right]$ \\[0.2cm]
    \hline \\
\end{tabular}
\caption{All leading order non-relativistic interaction operators.}\label{nonreloperatorstable}
\end{center}
\end{table}

Interactions between WIMPs and nuclei can be described in the framework of a non-relativistic effective field theory of WIMP-nucleon interactions. The possible quantum operators that describe such interactions are restricted by Galilean symmetry and can only be constructed as a combination of these five Hermitian operators:

\begin{equation}
    \mathbbm{1}_{\chi N},  \quad\qquad
    \mathbf{\hat{S}}_\chi,  \quad\qquad
    \mathbf{\hat{S}}_N,  \quad\qquad
    i\mathbf{\hat{q}},  \quad\qquad
    \mathbf{\hat{v}}^\perp,
\end{equation}
where index $\chi$ ($N$) refers to a WIMP (nucleon), $\mathbf{S}$ denotes a spin vector, $\mathbf{q}$ is the transferred momentum of the collision, and $\mathbf{v}^\perp\equiv \mathbf{v} + \mathbf{q}/2\mu_N$ is the transverse velocity as given by collisional velocity $\mathbf{v}$ and the reduced mass of the WIMP-nucleon system $\mu_N$.

Following the convention as established by \cite{Fitzpatrick13,Fitzpatrick14}, in table \ref{nonreloperatorstable} I have listed all linearly independent leading order operators that can be constructed from these building blocks, constrained by assuming a force mediating heavy particle of spin 1 or less. Because there are two types of nucleons, the parameter space of possible WIMP-nuclei interactions are doubled to a total number of 28.

The WIMP-nucleon Hamiltonian density can be written in a basis of isospin, represented by an upper index $\tau$, which is 0 for isoscalar and 1 for isovector coupling. In the former, WIMPs scatter off of all nucleons the same way; in the latter, the scattering off of protons and neutrons have opposite signs. By labeling each nucleon with an index $i$, the total Hamiltonian density for a nucleus of mass number $A$ can be written

\begin{equation}\label{hamiltoniandensitynucleons1}
    \hat{\mathcal{H}}(\mathbf{r})=\sum_{i=1}^A\sum_{\tau=0,1}\sum_{k=1}^{15}c_k^\tau\hat{\mathcal{O}}_k^{(i)}(\mathbf{r})t^\tau_{(i)},
\end{equation}
where $t^\tau_{(i)}$ is a matrix that projects a nucleon state onto an isospin basis. In the basis of proton and neutron couplings, the coupling coefficients are related like $c_k^p=(c_k^0+c_k^1)/2$ and $c_k^n=(c_k^0-c_k^1)/2$.

As has been shown in great detail in other sources \cite{Fitzpatrick13,Fitzpatrick14,CatenaSchwabe15}, this leads to a differential WIMP-nucleus cross section given by

\begin{align}
    \frac{\mathd \sigma(E_r,w^2)}{\mathd E_r} = & \frac{1}{2J_T+1}\frac{2m_T}{w^2}\sum_{\tau,\tau'}\Bigg\{\sum_{k=M,\Sigma'',\Sigma'}
    R_k^{\tau\tau'}\left( v_T^{\perp2},\frac{q^2}{m_N^2} \right) W_k^{\tau\tau'}(q^2)+ \nonumber\\
    & + \frac{q^2}{m_N^2}\Bigg(
    \sum_{k=\Phi'',\Phi''M,\tilde{\Phi}',\Delta,\Delta\Sigma'}
    R_k^{\tau\tau'}\left( v_T^{\perp2},\frac{q^2}{m_N^2} \right) W_k^{\tau\tau'}(q^2)\Bigg)\Bigg\},\label{eq:differentialcross}
\end{align}
where $E_r$ is the recoil energy in the rest frame of the target nucleus, $w$ is the collisional velocity, $J_T$ is the nucleus' spin, and $m_T$ ($m_N$) is the nucleus (nucleon) mass. The quantities denoted $R^{\tau\tau'}_k$ are the WIMP response functions, as found in appendix \ref{appendixA}. The quantities $W^{\tau\tau'}_k$ are the nuclear response functions, different for each nuclear isotope. Nuclear response functions for the 16 most abundant isotopes in the Sun are taken from \cite{CatenaSchwabe15}, where they have been calculated from ground state one-body density matrix elements. These 16 isotopes are, in order of mass abundance: H, $^4$He, $^{16}$O, $^{12}$C, $^{20}$Ne, $^{14}$N, $^{56}$Fe, $^{28}$Si, $^{24}$Mg, $^{32}$S, $^3$He, $^{59}$Ni, $^{40}$Ar, $^{40}$Ca, $^{27}$Al, $^{23}$Na.
\section{WIMP capture by the Sun}\label{section:capture}
In this section I present a theoretical background for dark matter accumulation in the Sun. In subsection \ref{ss:capture}, I review the theory contingent on the assumptions of a cold Sun and instant thermalization to a thermal density profile. In subsection \ref{ss:thermalization}, I relax the assumption of a cold Sun and present the theory by which I explore the thermalization process. In subsection \ref{ss:self-interaction}, I provide some remarks about capture and thermalization in the case of significant WIMP self-interaction.

\subsection{Capture and annihilation}\label{ss:capture}
WIMPs traveling through the Sun can collide with atomic nuclei in the solar interior. In some of these interactions the WIMP loses enough kinetic energy to be bound in orbit. The probablity per unit time for a WIMP to scatter to less than local escape velocity $v(r)$ is \cite{CatenaSchwabe15,Gould87}

\begin{equation}\label{capturesigma}
    \Omega_v^-(w) =
    \sum_T n_T w\,\Theta\left( \frac{\mu_T}{\mu_{+,T}^2}-\frac{u^2}{w^2} \right)
    \int_{E_k u^2/w^2}^{E_k \mu_T/\mu_{+,T}^2}\mathd E_r\frac{\mathd \sigma(E_r,w^2)}{\mathd E_r},
\end{equation}
where $m_\chi$ ($m_T$) is the dark matter (target nucleus) mass, $n_T$ is the target species number density, $u$ and $w=\sqrt{u^2+v(r)^2}$ are the dark matter particle velocities at point of scatter and at infinite radius, and $\mathd \sigma/\mathd E_r$ is the differential cross section, as given by equation \eqref{eq:differentialcross}. The lower bound of the integral represents the minimal energy transfer necessary for capture, while the upper limit is the highest possible energy transfer in an elastic collision, given by the WIMP's kinetic energy $E_k=m_\chi w^2/2$, and dimensionless parameters $\mu_T=m_\chi/m_T$ and $\mu_{+,T}=(\mu_T+1)/2$. The Heaviside function, $\Theta$, ensures that capture is kinematically possible.

Because interactions are weak and the Sun is optically thin, the WIMP capture rate by atomic nuclei per volume is given by

\begin{equation}\label{dCdV}
    \frac{\mathd C_c}{\mathd V} = \int_0^\infty \mathd u\frac{f(u)}{u}w\Omega_v^-(w),
\end{equation}
where $f(u)$ is the WIMP halo velocity distribution. Integrating over the full volume of the Sun gives the total capture rate

\begin{equation}\label{C}
    C_c =\int_0^{R_\odot}4\pi r^2 \frac{\mathd C_c}{\mathd V} \mathd r,
\end{equation}
where $R_\odot$ is the solar radius.

The other process which governs the amount of captured WIMPs is annihilation, which will come into effect when the concentration of WIMPs inside the Sun has become sufficiently high. A canonical value for the thermally averaged annihilation cross-section is $\langle\sigma_Av\rangle \simeq 3\cdot10^{-26}$ $\text{cm}^3\text{s}^{-1}$. However, recent studies have made more precise evaluations of this value \cite{SteigmanDasguptaBeacom12}. Following these results, the value used in this article is $\langle\sigma_Av\rangle = 2\cdot10^{-26}$ $\text{cm}^3\text{s}^{-1}$.

The total number of WIMPs annihilated per unit time is $C_aN^2$, where $N$ is the total number of trapped WIMPs and $C_a$ is an annihilation factor. The latter is given by

\begin{equation}\label{annihilationrate1}
    C_a=
    \frac{4\pi\langle\sigma_Av\rangle}{N^2}
    \int_0^{R_\odot} \epsilon^2(r)r^2 \mathd r,
\end{equation}
where $\epsilon(r)$ is the WIMP number density function. It is commonly assumed that the WIMPs thermalize to core temparature $T_c$ and follow a thermal profile,

\begin{equation}\label{thdensity}
    \epsilon(r) \propto \exp\left(-\frac{m_\chi \phi(r)}{k_B T_c}\right),
\end{equation}
where $\phi(r)$ is the gravitational potential. Because the Sun's core has a more or less constant density, the annihilation factor follows very closely the proportionality relation

\begin{equation}
    C_a\propto \langle\sigma_Av\rangle\, m_\chi^{3/2}.
\end{equation}

Assuming an instantaneous thermalization, the amount of WIMPs trapped within the Sun, $N$, is described by the following differential equation,

\begin{align}\label{capturediffeq}
    \frac{\mathd N}{\mathd t}=C_c-C_aN^2.
\end{align}
It has solution

\begin{align}\label{eq:Nsolution}
    N(t) = \sqrt{\frac{C_c}{C_a}} \tanh\left(\sqrt{C_cC_a}t\right),
\end{align}
which approaches an equilibrium solution $N_{eq}=\sqrt{C_c/C_a}$ as $t\rightarrow\infty$. The number of annihilation events per unit time follows the form

\begin{align}\label{eq:Gamma}
    \Gamma (t) = \frac{1}{2}C_a N^2(t) = \frac{1}{2}C_c\tanh^2\left(\sqrt{C_cC_a}t\right),
\end{align}
where the factor 1/2 comes from the fact that every annihilation event involves a WIMP pair.

\subsection{WIMP trajectories}\label{ss:thermalization}

The aim of this article is to study the process of thermalization, to evaluate the thermalization time scale and eventual density profile, for different types of WIMP-nucleon interactions. This is done by Monte-Carlo integration, by following WIMP trajectories from the first scattering event that binds a WIMP to the Sun's gravitational field, to down-scattering to orbits that are in thermal equilibrium with the Sun's core. In order to accurately randomize these trajectories, I evaluate the probability density functions and 3-dimensional kinematics that govern this process.

A WIMP's orbit in or around a spherically symmetric massive body is completely described by its total energy $E$ and angular momentum $J$. The innermost and outermost radii of a WIMP's orbit fulfill the relation

\begin{equation}
    E=m_\chi\phi(r)+\frac{J^2}{2m_\chi r^2},
\end{equation}
where $\phi(r)$ is the gravitational potential. The distance traveled through a shell of thickness $\mathd r$, per orbital period, is

\begin{equation}
    \mathd s=2 \mathd r \left( 1-(\frac{J}{m_\chi rw})^2 \right)^{-1/2},
\end{equation}
where the factor $2$ is due to the fact that a WIMP travels through a shell twice per orbital period. The time it takes to pass through this shell is $\mathd t = w^{-1}\; \mathd s$, which gives the orbital time by integration from minimal to maximal radius.

The WIMP-nucleus interactions are weak and the Sun is optically thin. For the case of a cold Sun, neglecting any thermal motion of the target nuclei, the probability of scatter in a thin shell of thickness $\mathd r$ during one orbital period is

\begin{equation}\label{scprobcold}
    \mathd P_{sc} = \mathd s\; n_T \int_{0}^{E_k(w) \mu_T/\mu_{+,T}^2}\mathd E_r\frac{\mathd \sigma(E_r,w^2)}{\mathd E_r}.
\end{equation}
By including the thermal motion of the target nuclei, the probability of scatter per orbital period becomes

\begin{equation}\label{scprobwarm}
    \mathd P_{sc} = \frac{\mathd s\; n_T}{w}\int_{-\infty}^{\infty}\mathd \tilde{w}\int_{0}^{E_k(\tilde{w}) \mu_T/\mu_{+,T}^2}\mathd E_r\;
    \frac{\mathd \sigma(E_r,\tilde{w}^2)}{\mathd E_r}\;f_T(\tilde{w})\,\tilde{w},
\end{equation}
where $\tilde{w}$ is the collisional velocity between the WIMP and target nucleus, and $f_T(\tilde{w})$ is the velocity distribution of a nucleus in the WIMP rest frame, assumed to be a thermal Maxwell-Boltzmann distribution boosted by velocity $w$ and normalized to 1. The factor $\mathd s/w$ corresponds to the time spent in the shell, while the factor $f_T(\tilde{w})\,\tilde{w}$ in the integrand accounts for the number of encounters of a certain collisional velocity. In the limit $w\gg k_BT/m_T$, $f_T(\tilde{w})$ becomes a narrow function peaked around $\tilde{w}=w$, such that equation \eqref{scprobwarm} becomes equivalent to \eqref{scprobcold}. Integrating $\mathd P_{sc}$ over all radii gives the total scattering probability per orbital period and, given the time of such a period, the average time spent on this orbit. Expressing $\mathd P_{sc}$ as a probability density function over $r$ allows for a randomization of the scattering radius.

Given a scattering event where the collisional velocity $\tilde{w}$ and the WIMP velocity $w$ are known, the thermal velocity of the nuclei, $v_{th}$, is not unique. Its component along the WIMP trajectory axis (in the solar rest frame), $v_{thz}$, can be anything in range $w-\tilde{w} \le v_{thz} \le w+\tilde{w}$. The perpendicular component of the thermal velocity fulfills that $v_{th\perp}^2+(v_{thz}-w)^2=\tilde{w}^2$. Expressing this condition as a delta function over $\tilde{w}$ and integrating over all other variables gives the probability density function for $v_{thz}$:

\begin{gather}
    \tilde{f}_T(v_{thz})\, \mathd v_{thz} = \nonumber \\
    = \int \mathd\theta\, v_{th\perp} \mathd v_{th\perp} \delta\left(\tilde{w}-\sqrt{(v_{thz}-w)^2+v_{th\perp}^2}\right)
    \left( \frac{m_T}{2\pi k_BT} \right)^{3/2}
    \exp \left( -\frac{m_T}{2k_BT}v_{th}^2\right)\mathd v_{thz} = \nonumber \\ 
    = 2\pi \tilde{w} \left( \frac{m_T}{2\pi k_BT} \right)^{3/2} \exp \left( -\frac{m_T}{2k_BT}(\tilde{w}^2-w^2+2wv_{thz}) \right)\mathd v_{thz},
\end{gather}
where there is one factor $v_{th\perp}$ from the cylindrical coordinate system Jacobian and one factor $\tilde{w}/v_{th\perp}$ from the inner derivative of the delta function. Integrating this function over the range of possible $v_{thz}$ gives the boosted Maxwell-Boltzmann distribution. After finding the collisional velocity, thermal velocity and recoil energy of a collision, there are two remaining degrees of freedom. They are found in the angular orientation of $v_{th\perp}$ and in one scattering angle, both of which are trivially randomized in the rest frame of the target nucleus where all angles are equiprobable.

Thus I have all the probability density functions necessary in order to randomize the scattering target, scattering radius, collisional velocity, target thermal velocity, recoil energy, scattering angles, and subsequent WIMP energy and angular momentum of the new orbit.

The very first scattering event, when a WIMP from the halo scatters and becomes bound in orbit, is randomized in almost the same manner. In this case, the radius of scattering is given by the integrand in equation \eqref{C}, and the WIMP velocity by the integrand in equation \eqref{dCdV}. The angular momentum after first scattering is trivially randomized, given by the fact that all incoming solid angles are equiprobable, due to the homogeneity of the WIMP halo distribution and spherical symmetry of the system.

I do not concern myself with the influence of planets and how they affect WIMP trajectories, as this is beyond the scope of this article. This issue is non-trivial and has been discussed for decades, with many twist and turns with regards to the evaluated capture rates \cite{Lundberg04,Gould91,Damour99,Gould01,Peter09a,Peter09b,Peter09c,Sivertsson12}. Most recently, is has been argued that completely neglecting the planets and regarding the Sun as being in free space is a fair approximation \cite{Sivertsson12}, both in terms of the total capture rate and the thermalization process itself.

\subsection{A note on WIMP self-interaction}\label{ss:self-interaction}

It has been pointed out that WIMP self-interaction can increase the capture rate and amplify the resulting neutrino signal \cite{Zentner09}, as an already amassed concentration of trapped WIMPs will itself constitute a scattering target for WIMPs from the galactic halo. It has been demonstrated that, within current limits to the WIMP self-interaction cross section, such effects can amplify the neutrino signal by several orders of magnitude \cite{CatenaWidmark16}.

In this case, the differential equation \eqref{capturediffeq} that describes how the number of captured WIMPs change over time is modified to read

\begin{align}\label{capturediffeqself}
    \frac{\mathd N}{\mathd t}=C_c-C_aN^2+C_sN,
\end{align}
where the last term corresponds to the capture rate by WIMP self-interaction. The quantity $C_s$ has unit of inverse time, such that $C_s^{-1}$ corresponds to the average time it takes before a thermalized WIMP interacts with a WIMP from the galactic halo. (There are some minor and negligable corrections to this statement, as in some cases a collision can result in one of the two WIMPs being ejected, resulting in no net gain or loss of captured WIMPs; in extremely rare cases both WIMPs can be ejected, resulting in evaporation.) The total energy that is distributed between the two WIMPs in the collisions is of order $m_\chi v^2_{esc}(r=0)/2$. As collisions between WIMPs are resonant, very long orbits are unlikely and need not be taken into account. Rather, the two WIMPs involved in such a capture by self-interaction will spend most of their thermalization time at small radii. If that thermalization time is longer or of about equal value with $C_s^{-1}$, the assumption of instant thermalization is broken. The kinetic energy of the WIMPs would not dissipate quickly enough to counteract the energy input from the halo, resulting in a heating up of the density profile of thermalized WIMPs. Given time, an equilibrium between capture and annihilation would still be reached, although with a lower annihilation coefficient $C_a$ and a higher total concentration of trapped WIMPs.

In the calculations of neutrino signal amplification due to WIMP self-interaction \cite{CatenaWidmark16}, the self-interaction cross-section is limited by N-body simulations \cite{Rocha13} to an approximate upper value of
\begin{equation}
    \sigma_{\chi\chi} < 0.1\;\frac{m_\chi}{\text{g}}\;\text{cm}^2 = 1.78\times10^{-25}\; \frac{m_\chi}{\text{GeV}}\;\text{cm}^2.
\end{equation}
This corresponds to a value of $C_s^{-1}=6.5\times10^{8}$ years. If the thermalization time, starting from a mid-range energy of order $\sim m_\chi v^2_{esc}(r=0)/4$, is longer than this time scale, then WIMPs will not be able to re-thermalize before being subject to another halo WIMP interaction.
\section{Results}\label{section:results}

The thermalization process has been simulated by Monte-Carlo integration; by sampling a large number of WIMP trajectories I have calculated thermalization time scales and subsequent thermal density profiles. I have considered a WIMP in the mass range of 10--1000 GeV and spin 1/2. The galactic WIMP halo is assumed to follow a Maxwell-Boltzmann distribution with velocity dispersion of 270 km/s, a Local Standard of Rest velocity of 220 km/s, and a local dark matter density of 0.4 GeV/cm$^3$. Solar densities and temperatures are taken from \cite{Bahcall01}, as used also in \cite{Gondolo04}. The atomic nuclei of the Sun are assumed to follow a Maxwell-Boltzmann velocity distribution, given by the local solar temperature.

\subsection{Density profiles}

It is commonly assumed that thermalized WIMPs follow a thermal profile, as described by equation \eqref{thdensity}. This assumption has been tested by following a single WIMP's trajectory, starting from a state of thermal equilibrium, as it collides with atomic nuclei within the solar medium. I have considered all WIMP-nucleon interaction operators of table \ref{nonreloperatorstable}, in both isoscalar and isovector couplings, with WIMP masses of 10, 100 and 1000 GeV.

\begin{figure}[tbp]
\centering
\includegraphics[width=.9\textwidth,origin=c]{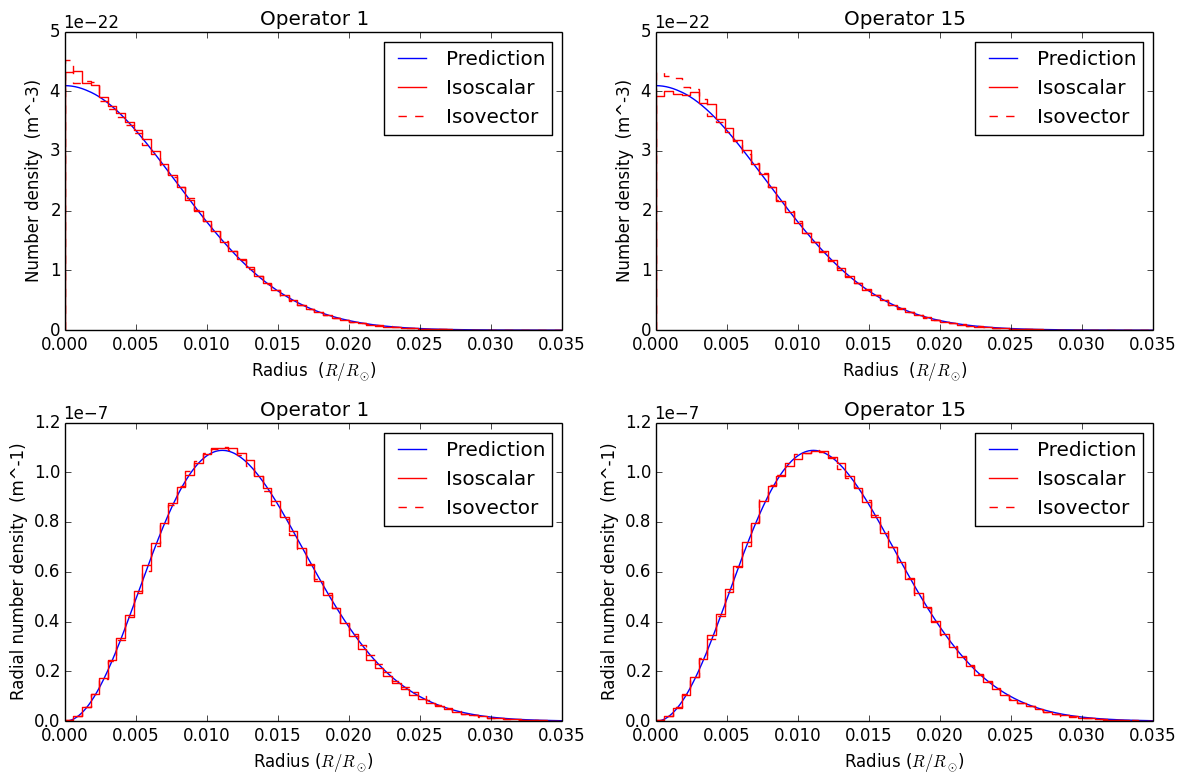}
\caption{\label{fig:profile} Thermalization profiles for a WIMP of mass 100 GeV, with WIMP-nucleon interaction operators $\hat{\mathcal{O}}_1$ (left) and $\hat{\mathcal{O}}_{15}$ (right), with isoscalar (solid red) and isovector (dashed red) coupling. The thermal profile (solid blue) is the commonly assumed distribution, given by the Sun's core temperature and gravitational potential. The upper panels show the time averaged number density as a function of radius, normalized to unity. The lower panels show the probability density for a WIMP to be located at radius $R$ at a given point in time.}
\end{figure}

Thermalization profiles for operators $\hat{\mathcal{O}}_1$ and $\hat{\mathcal{O}}_{15}$ are shown in figure \ref{fig:profile}. Although these two operators are of different nature, the density profiles are practically the same. The upper panels of figure \ref{fig:profile} display the number density as a function of radius; the innermost bins are somewhat noisy, explained by the fact that these bins represent very small volumes. An alternative representation of the distribution is visible in the lower panels, which display the probability that a WIMP is located at a certain radius. The time averaged total energy of the thermalized WIMPs takes the approximate value $\langle E \rangle \simeq 3k_BT$, which fits well with the fact that a particle in the potential well of the Sun has 6 degrees of freedom. Note that the average energy is higher than the median due to the high-energy tail of the distribution. For a given point in time, most of the thermalized WIMPs have energies below $\sim 2.2k_BT$.

I have found that all operators tend to the same profile, very accurately described by the standard assumption of a thermal profile. The resulting value for the annihilation coefficient $C_a$, given by equation \eqref{annihilationrate1}, differs at most by a few percent.

\subsection{Thermalization time scales}

Thermalization time scales are calculated by sampling a large number of WIMP trajectories, where each trajectory begins with a first scattering event that binds a WIMP from the galactic halo to the Sun's gravitational field, followed by down-scattering to lower energies. The trajectory ends when the WIMP can be considered thermalized, chosen as the first time that its energy goes below the time-averaged energy of the thermal distribution $\langle E \rangle \simeq 3k_BT$. I present my results as a time median of these trajectories (the mean value is not very illustrative due to orbit outliers of very long radii).

\begin{figure}[tbp]
\centering
\includegraphics[width=.7\textwidth,origin=c]{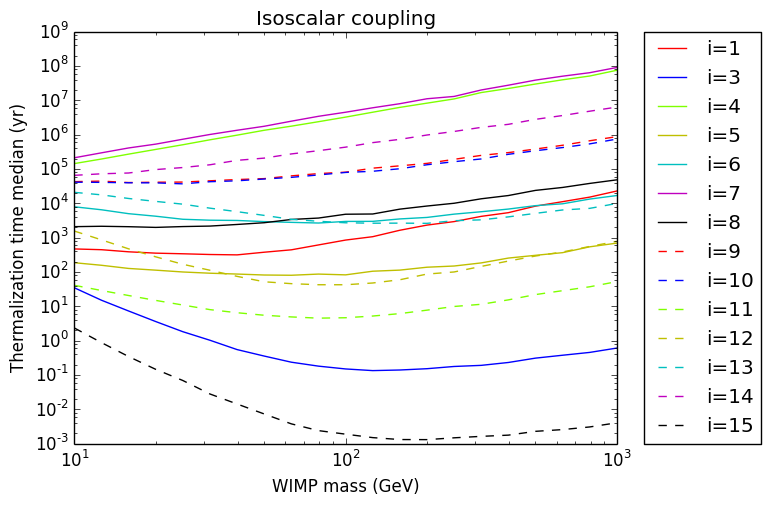}
\includegraphics[width=.7\textwidth,origin=c]{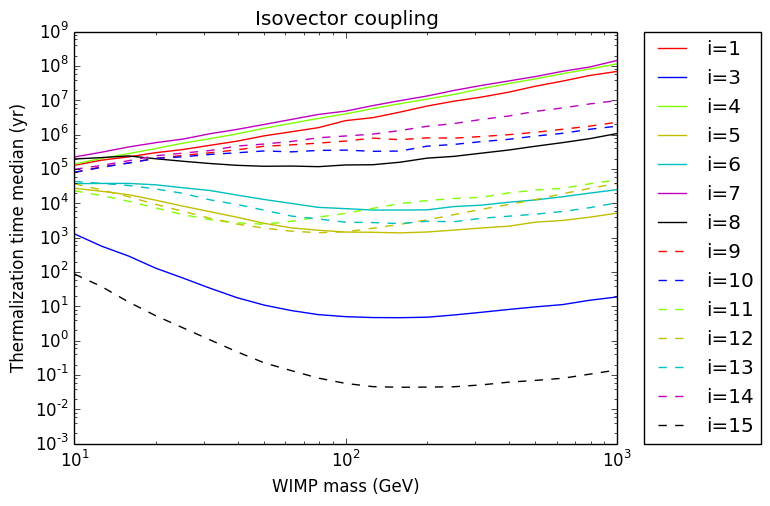}
\caption{\label{fig:sigmanorm} Thermalization time scales for operators $\hat{\mathcal{O}}_i$, with isoscalar (upper panel) and isovector (lower panel) couplings. The respective coupling constants are set such that the WIMP-nucleon cross section at collisional velocity 1000 km/s has value $\sigma_{\chi N}=10^{-44}$ cm$^2$.}
\end{figure}

\begin{figure}[tbp]
\centering
\includegraphics[width=.7\textwidth,origin=c]{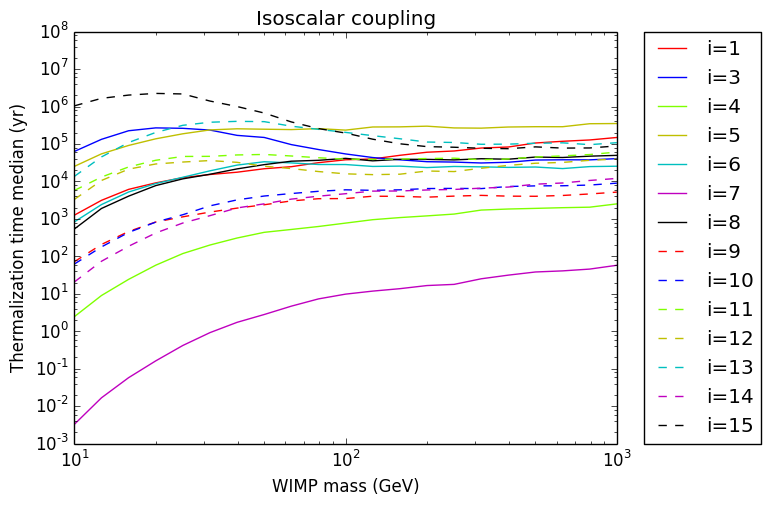}
\includegraphics[width=.7\textwidth,origin=c]{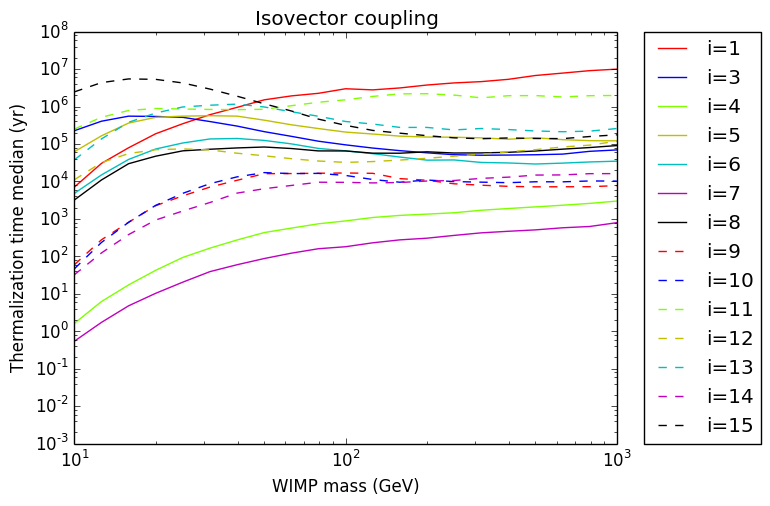}
\caption{\label{fig:LUXlimit} Thermalization time scales for operators $\hat{\mathcal{O}}_i$, with isoscalar (upper panel) and isovector (lower panel) couplings. The respective coupling constants are set to the limit given by the LUX direct detection experiment. This sets a lower limit to the thermalization time.}
\end{figure}

\begin{figure}[tbp]
\centering
\includegraphics[width=.7\textwidth,origin=c]{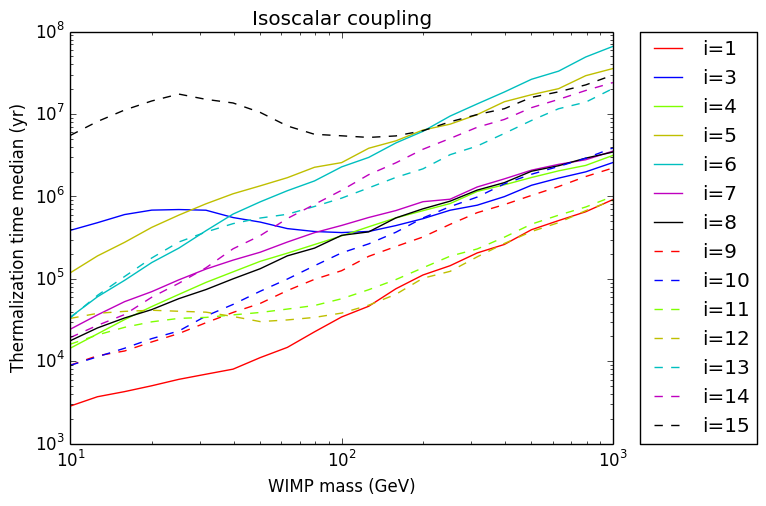}
\includegraphics[width=.7\textwidth,origin=c]{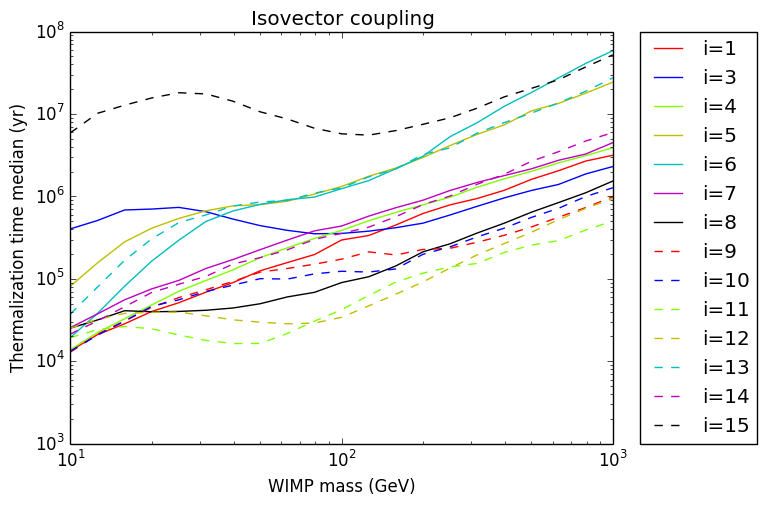}
\caption{\label{fig:equinorm} Thermalization time scales for operators $\hat{\mathcal{O}}_i$, with isoscalar (upper panel) and isovector (lower panel) couplings. The respective coupling constants are normalized to a specific capture rate, $C_c=1/(C_a t_\odot^2)$, for which the amount of captured WIMPs is close to equilibrium and significant annihilation has come into effect.}
\end{figure}

These results are presented using three different normalizations. In figure \ref{fig:sigmanorm}, the WIMP-nucleon cross section is normalized to a specific value. While this serves a purpose of record keeping, it does not illustrate very well the greater picture when it comes to the capture rate and expected neutrino signal. Furthermore, for a lot of operators direct detection experiments have excluded such large cross sections. This is shown in figure \ref{fig:LUXlimit}, where the coupling constants are set to the limits provided by the Large Underground Xenon (LUX) direct detection experiment. In figure \ref{fig:equinorm}, the interaction strength is normalized such that all operators give rise to the same total capture rate. In this manner, it is possible to relate the thermalization time scales to not only to the capture rate, but also to the rate of annihilation and resulting flux of neutrinos.

The thermalization time medians for all operators $\hat{\mathcal{O}}_{i}$ are visible in figure \ref{fig:sigmanorm}, where the coupling constants are normalized to values such that the WIMP-nucleon cross section at collisional velocity 1000 km/s is $\sigma_{\chi N}=10^{-44}$ cm$^2$. The cross section and thermalization time have an inverse proportionality. The longest thermalization times in this figure are from operators that scatter almost exclusively on hydrogen, while other operators also interact with heavier nuclei. A lot of operators with isovector couplings have significantly longer time scales than their isoscalar counterparts, which is due to destructive interference between proton and neutron scattering. Although the general feature is that the thermalization time increases with WIMP mass, some operators have their shortest thermalization time for a mid-range WIMP. This is due to resonant effects; for example, operator $\hat{\mathcal{O}}_{15}$ (isoscalar and isovector) scatters predominantly off of $^{56}$Fe and has its shortest thermalization time for a corresponding WIMP mass. Depending on the governing operator, the 90th percentile to the thermalization time is a factor 1.5--10 larger than the median, where this factor is strongly correlated with the fraction of time spent on long orbits. Such behavior, of spending a majority of time on long orbits, is exhibited by operators $\hat{\mathcal{O}}_4$, $\hat{\mathcal{O}}_7$ and the isovector component of $\hat{\mathcal{O}}_1$, and in the higher mass range also $\hat{\mathcal{O}}_{11}$, $\hat{\mathcal{O}}_{12}$ and the isoscalar component of $\hat{\mathcal{O}}_1$. In the remainder of parameter space, however, the majority of the thermalization time is actually spent in the solar interior. This behavior is especially pronounced for operators with a strong dependence on transferred momentum, such as $\hat{\mathcal{O}}_6$ and $\hat{\mathcal{O}}_{15}$. For these operators, the WIMPs down-scatter to orbits within the solar interior very quickly, but on the other hand the cross section drops dramatically with lower collisional velocities, resulting in a very slow energy loss in the very end of the thermalization process.

In figure \ref{fig:LUXlimit}, the coupling constants are set to the limit provided by the LUX direct detection experiment. These limits were calculated in \cite{Catena15}, using the first publication of results from LUX \cite{Akerib14}. Since then, stronger limits have been provided, most recently with \cite{Akerib17}. The experiment's sensitivity with respect to the WIMP-nucleon cross section has increased by about a factor 5. They have also seen a downward fluctuation in their background signal, which have put an even stricter limit to the cross section in the higher WIMP mass range (an improvement of about a factor 8 in the 90 \% C.L. between the old and new source). For the sake of simplicity, I use the coupling constant limits provided by \cite{Catena15}, but increase the time medians by a factor 5 to account for the sensitivity difference between the old and new LUX limits. There is a very large spread also in this figure, due to the varying quality of the coupling constant limits for the different interaction operators. The limits are contingent on the nuclear structure and evaluated nuclear response function of xenon (the LUX detector medium), as well as the current local WIMP halo density. All thermalization time scales are significantly shorter than the age of the Sun, $t_\odot \simeq 4.5\times10^9$ years, so there is still a large margin before direct detection experiments have excluded negligible thermalization times.

In most scenarios where there is hope of detecting a high-energy neutrino signal coming from the Sun, the capture rate must be high enough for annihilation to have come into significant effect. Because the annihilation rate and the neutrino signal is proportional to the WIMP density squared, as is expressed in equation \eqref{capturediffeq}, significant annihilation presupposes that the number of trapped WIMPs is close to its equilibrium amount. In figure \ref{fig:equinorm}, the thermalization time medians are presented for operator coupling constants that are normalized such that the total capture rate is $C_c=1/(C_a t_\odot^2)$, where $t_\odot$ is the age of the Sun. By using this value for $C_c$, the number of annihilation events per unit time, as given by equation \eqref{eq:Gamma}, is $\tanh^2(1)\simeq58 \%$ of its equilibrium value. The values for the capture rates that have been calculated in this project are in accordance with the results of \cite{CatenaSchwabe15}. With this normalization a new picture emerges. The thermalization time for a specific WIMP mass differs at most $\sim3$ orders of magnitude between operators. The longest thermalization time scales are not for operators that scatter against hydrogen into very long orbits (primarily $\hat{\mathcal{O}}_4$ and $\hat{\mathcal{O}}_7$), but rather for operators with a strong dependence on transferred momentum ($\hat{\mathcal{O}}_6$ and $\hat{\mathcal{O}}_{15}$).

For a few operators, a comparison between the last two figures shows that a close to equilibrium amount of trapped WIMPs is excluded by LUX, as the lower limit to the thermalization time in figure \ref{fig:LUXlimit} is higher than the value presented in figure \ref{fig:equinorm}. This is the case for isoscalar component of operators $\hat{\mathcal{O}}_1$, $\hat{\mathcal{O}}_{11}$ and $\hat{\mathcal{O}}_{13}$, and the isovector component of operators
$\hat{\mathcal{O}}_1$, $\hat{\mathcal{O}}_5$, $\hat{\mathcal{O}}_8$, $\hat{\mathcal{O}}_{11}$,
$\hat{\mathcal{O}}_{12}$ and $\hat{\mathcal{O}}_{13}$. Most of them are excluded by a small margin and only in the lower mass range, but for example the values of $\hat{\mathcal{O}}_{11}$ with isovector coupling differ by almost 2 orders of magnitude.
\section{Conclusions}\label{section:conclusions}

I have studied the thermalization process of WIMP capture by the Sun. I have considered a WIMP in the mass range of 10--1000 GeV, spin 1/2, and an interaction with atomic nuclei described by non-relativistic effective field theory with 28 degrees of freedom.

I have found that the density profiles of thermalized WIMPs agree very well with the standard assumption of a thermal profile. The thermalization time, on the other hand, varies dramatically depending on what operator that dominates the interaction. Using limits provided by the LUX direct detection experiment, most operators have a very large margin before the assumption of instantaneous thermalization breaks down.

In order to detect a neutrino signal coming from WIMP annihilation in the Sun, the rate of annihilation must be sufficiently high. In most scenarios for which there is hope of detecting such a signal, the number of captured WIMPs must at the very least be close to its equilibrium solution. By normalizing the coupling strength such that $C_c=1/(C_a t_\odot^2)$, for which the neutrino flux is 58~\% of its value at equilibrium, I find thermalization time medians in the approximate range of $10^4$--$10^8$ years. Compared to the 4.5 billion year life time of the solar system, these time scales are short. In other words, if the capture rate is sufficiently large to give rise to an equilibrium (or almost equilibrium) number of trapped WIMPs at present day, then the assumption of instantaneous thermalization is valid. However, making the cross section one order of magnitude weaker could already be problematic in some cases, especially for WIMPs in the higher mass range.

In terms of the thermalization process, the effective field theory operators of table \ref{nonreloperatorstable} can be categorized into two groups. In the first group, WIMPs spend most of their thermalization time on their first few orbits; operators that scatter almost exclusively off of hydrogen exhibit this behavior. In the second group, WIMPs spend most of their thermalization time on short orbits in the solar interior, a behavior that is especially pronounced for operators with quadratic or cubic dependence on transferred momentum, which is due to the decreasing WIMP velocity and rate of interaction. It must be noted that in my analysis I have assumed the interactions to be dominated by only one operator, while in reality a combination of operators is expected. This is especially relevant for operators with a strong dependence on collisional velocity. For example, operator $\hat{\mathcal{O}}_{15}$ might dominate the first few scattering events and thus the probability for a WIMP to be captured, but as the collisional velocity decreases operator $\hat{\mathcal{O}}_1$ could start to dominate. Such a behavior would serve to hasten the thermalization process, as a higher interaction rate unequivocally makes the WIMP lose its energy quicker.

As shown in \cite{CatenaWidmark16}, WIMP self-interaction could potentially amplify the capture rate and resulting neutrino signal, especially so if the capture by nuclei is insufficient in terms of creating an equilibrium at present time. In such a scenario, WIMP self-interaction could increase the capture rate to the extent that equilibrium is reached anyway. In fine-tuned cases of WIMP-nuclei interactions with quadratic or cubic dependence on transferred momentum, the thermalization time in the solar interior could be longer than the average time between self-interactions with halo WIMPs. This would result in a heating up of the density profile, an effect that suppresses the annihilation rate and the resulting neutrino signal.

As is mentioned in section \ref{section:introduction}, the thermalization time is especially relevant if the solar system travels through substructures in the galactic WIMP halo, which would give rise to a time-varying capture rate. In \cite{Koushiappas09} they consider the effect of passing through halo substructures with local over-densities of 2 and even 3 orders of magnitude, crossing-times in range of $10^2$--$10^7$ years, and mean time between encounters of $10^6$ years and upwards. While the capture rate is proportional to the local WIMP density, the response in annihilation and neutrino flux is dependent on the thermalization process. If the thermalization time is longer than the time scales of the density fluctuations, the response of the neutrino signal will be shifted and stretched. Given the results presented in this article, this can clearly be the case.

In summary, the thermalization time and general behavior differs greatly between different types of WIMP-nuclei interactions. For some operators most of the thermalization process is spent on very long orbits; for other operators the very opposite is the case. Either way, the standard assumption of instant thermalization is valid in most cases of interest in the effective field theory framework, although not necessarily by a large margin.

\newpage
\appendix
\section{Dark matter response functions}\label{appendixA}
Below are the dark matter response functions, as found in equation \eqref{eq:differentialcross}.

\begin{eqnarray*}
R_M^{\tau\tau'} &=& c_1^\tau c_1^{\tau'}+\frac{j_\chi(j_\chi+1)}{3}\left( \frac{q^2}{m_N^2}v_T^{\perp2}c_5^\tau c_5^{\tau'}+v_T^{\perp2}c_8^\tau c_8^{\tau'}+\frac{q^2}{m_N^2}c_{11}^\tau c_{11}^{\tau'} \right)        \\
R_{\Phi''}^{\tau\tau'} &=& \frac{q^2}{4m_N^2}c_3^\tau c_3^{\tau'}+\frac{j_\chi(j_\chi+1)}{12}\left( c_{12}^\tau-\frac{q^2}{m_N^2}c_{15}^\tau \right)\left( c_{12}^{\tau'}-\frac{q^2}{m_N^2}c_{15}^{\tau'} \right)        \\
R_{\Phi''M}^{\tau\tau'} &=& c_3^\tau c_1^{\tau'}+\frac{j_\chi(j_\chi+1)}{3}\left( c_{12}^\tau-\frac{q^2}{m_N^2}c_{15}^\tau \right)c_{11}^{\tau'}        \\
R_{\tilde{\Phi}'}^{\tau\tau'} &=& \frac{j_\chi(j_\chi+1)}{12}\left( c_{12}^\tau c_{12}^{\tau'}+\frac{q^2}{m_N^2}c_{13}^\tau c_{13}^{\tau'} \right)        \\
R_{\Sigma''}^{\tau\tau'} &=& \frac{q^2}{4m_N^2}c_{10}^\tau c_{10}^{\tau'}+\frac{j_\chi(j_\chi+1)}{12}\Big[ c_4^\tau c_4^{\tau'}+\frac{q^2}{m_N^2}(c_4^\tau c_6^{\tau'}+c_6^\tau c_4^{\tau'})+        \\
& & +\frac{q^4}{m_N^4}c_6^\tau c_6^{\tau'}+v_T^{\perp2}c_{12}^\tau c_{12}^{\tau'}+\frac{q^2}{m_N^2}v_T^{\perp2}c_{14}^\tau c_{14}^{\tau'} \Big] \\
R_{\Sigma'}^{\tau\tau'} &=& \frac{1}{8}\left( \frac{q^2}{m_N^2}v_T^{\perp2}c_3^\tau c_3^{\tau'}+v_T^{\perp2}c_7^\tau c_7^{\tau'} \right)+\frac{j_\chi(j_\chi+1)}{12}\Big[c_4^\tau c_4^{\tau'}+ \frac{q^2}{m_N^2}c_9^\tau c_9^{\tau'} +        \\
& & +\frac{v_T^{\perp2}}{2}\left( c_{12}^\tau-\frac{q^2}{m_N^2}c_{15}^\tau \right)\left( c_{12}^{\tau'}-\frac{q^2}{m_N^2}c_{15}^{\tau'}+\frac{q^2}{2m_N^2}v_T^{\perp2}c_{14}^\tau c_{14}^{\tau'} \right)\Big] \\
R_{\Delta}^{\tau\tau'} &=& \frac{j_\chi(j_\chi+1)}{3}\left( \frac{q^2}{m_N^2}c_5^\tau c_5^{\tau'}+c_8^\tau c_8^{\tau'} \right)        \\
R_{\Delta\Sigma'}^{\tau\tau'} &=& \frac{j_\chi(j_\chi+1)}{3}\left( c_5^\tau c_4^{\tau'}-c_8^\tau c_9^{\tau'} \right)
\end{eqnarray*}

\acknowledgments

I would like to thank Joakim Edsj\"o and Sofia Sivertsson, who have provided valuable guidance, insightful discussion and cross-checking of results throughout the development of this project. I would also like to give thanks to Riccardo Catena, who first introduced me to the subject of dark matter capture by the Sun.

\end{document}